\documentclass[amsmath,amssymb]{revtex4}
\begin{document}
\title{Tables of the Appell Hypergeometric Functions $F_2$}

\author{Jonathan Murley}%
\author{Nasser Saad}
\email{Jmurley@stu1.cs.upei.ca}
\email{nsaad@upei.ca}
\affiliation{Department of Mathematics and Statistics, University of Prince Edward Island 
Charlottetown, Prince Edward Island C1A 4P3, Canada}%


\def\dbox#1{\hbox{\vrule  
                        \vbox{\hrule \vskip #1
                             \hbox{\hskip #1
                                 \vbox{\hsize=#1}%
                              \hskip #1}%
                         \vskip #1 \hrule}%
                      \vrule}}
\def\qed{\hfill \dbox{0.05true in}}  
\def\square{\dbox{0.02true in}} 
\begin{abstract}
\noindent{\bf Abstract:} The generalized hypergeometric function $_qF_p$ is a power series in which the ratio of successive terms is a rational function of the summation index.  The Gaussian hypergeometric functions $_2F_1$ and $_3F_2$ are most common special cases of the generalized hypergeometric function $_qF_p$. The Appell hypergeometric functions $F_q$, $q=1,2,3,4$ are  product of two hypergeometric functions $_2F_1$  that appear in many areas of mathematical physics. Here, we are interested in the Appell hypergeometric function $F_2$ which is known to have a double integral representation. As demonstrated by Opps, Saad, and Srivastava (J. Math. Anal. Appl. 302 (2005) 180-195), the double integral representation of $F_2$ can be reduced to a single integral that can be easily evaluated for certain values of the parameters in terms of $_2F_1$ and $_3F_2$. Using many of the reduction formulas of $_2F_1$ and $_3F_2$ and the representation of $F_2$ in terms of a single integral, we have begun to tabulate new reduction formulas for $F_2$.
\end{abstract}
\vskip0.1 true in
\maketitle
   \vskip0.1true in                        
\noindent{\bf keywords:} {~Multiple hypergeometric functions, Appell series, Generalized hypergeometric function, Gauss hypergeometric function,  Clausen hypergeometric function,  Reduction and transformation formulas, Radiation field, Hubbell integral.
}


\section{Introduction}	
\noindent  Appell hypergeometric functions $F_D,D=1,2,3,4$  play an important role in mathematical physics (\cite{app}-\cite{bailey}, \cite{Exton}, \cite{erdely1}-\cite{erdely2}, \cite{slater}, \cite{sri}). In particular, the Appell hypergeometric series $F_2$ arises frequently in various physical and chemical applications (\cite{beckert}, \cite{taraso1} to \cite{opps}, \cite{salman} to \cite{trasov2} ). The exact solution of number of problems in quantum mechanics has been given \cite{landau} in terms of Appell's function $F_2$.
It is defined by [\cite{slater}, p. 211, Equation (8.1.4)]: 
\begin{equation}\label{eq1}
F_2(\sigma,\alpha_1,\alpha_2;\beta_1,\beta_2;x,y)=\sum\limits_{m=0}^\infty\sum\limits_{n=0}^\infty {(\sigma)_{m+n}(\alpha_1)_m(\alpha_2)_n\over (\beta_1)_m(\beta_2)_n} \, {x^m \over m!} \, {y^n \over n!},
\end{equation}
for $|x| + |y| <1; \, \beta_j \in {\Bbb C} \backslash {\Bbb Z}_0^-; \, {\Bbb Z}_0^-:=\lbrace 0,-1,-2, \ldots \rbrace,$ and
$(\lambda)_k$ denotes  the Pochhammer symbol defined, in terms of Gamma functions, by
$$(\lambda)_k:= {\Gamma(\lambda+k)\over\Gamma(\lambda)}=\left\{ \begin{array}{ll}
 1 &\mbox{ if\quad $\left( k=0; \, \lambda \in {\Bbb C} \backslash \lbrace 0 \rbrace \right) $} \\
  \lambda(\lambda+1)(\lambda+2)\dots(\lambda+k-1) & \mbox{ if\quad $\left( k \in {\Bbb N}; \, \lambda \in {\Bbb C} \right)$}
       \end{array} \right.
$$
where $\Bbb N$ being the set of {\it positive} integers. Further, it also has the following double integral representation [\cite{slater}, p. 214, Equation (8.2.3)]:
\begin{eqnarray}\label{eq2}
F_2(\sigma,\alpha_1,\alpha_2;\beta_1,\beta_2;x,y)&=&{\Gamma(\beta_1)\Gamma(\beta_2)\over \Gamma(\alpha_1)\Gamma(\alpha_2)\Gamma(\beta_1-\alpha_1)\Gamma(\beta_2-\alpha_2)}\int_0^1
\int_0^1u^{\alpha_1-1}\tau^{\alpha_2-1}\nonumber\\
&~&\times(1-u)^{\beta_1-\alpha_1-1}(1-\tau)^{\beta_2-\alpha_2-1}(1-xu-y\tau)^{-\sigma}du \, d\tau
\end{eqnarray}
where ${\frak R}\left( \beta_j \right) > {\frak R} \left( \alpha_j \right) >0, \, j=1,2$ and $|x| +|y| <1$. Recently, Opps et al.  \cite{opps} used the Euler's integral representation of the Gauss hypergeometric function [\cite{slater}, p. 20, Equation (1.6.6)] 
\begin{equation}\label{eq3}
{}_2F_1\bigg({a,b \atop c} \bigg|z \bigg)={\Gamma(c)\over \Gamma(b)\Gamma(c-b)}
\int_0^1\tau^{b-1}(1-\tau)^{c-b-1}(1-z\tau)^{-a}d\tau 
\end{equation}
to reduce the double integral representation (\ref{eq2}) into a single integral in term of $_2F_1$, 
\begin{equation}\label{eq4}
F_2(\sigma;\alpha_1,\alpha_2;\beta_1,\beta_2;x,y)
={\Gamma(\beta_1)\over \Gamma(\alpha_1)\Gamma(\beta_1-\alpha_1)}
\int_0^1 {u^{\alpha_1-1}(1-u)^{\beta_1-\alpha_1-1}\over (1-xu)^{\sigma}}{}_2F_1\bigg({\sigma,\alpha_2 \atop \beta_2}\bigg|{y \over 1-xu}\bigg)~du
\end{equation}
for ${\frak R}\left( \beta_1 \right) > {\frak R}\left( \alpha_1 \right) >0$, and $|x| + |y| <1.$ Using some properties of $_2F_1$, they prove the following theorem [\cite{opps}, Theorem 1]:
\vskip0.1true in
\noindent {\bf Theorem 1.} {\it For $|x|+|y|<1$, the Appell hypergeometric function $F_2$ is given by
\begin{eqnarray}\label{eq5}
F_2(a+1;\alpha_1,1;\beta_1,2;x,y)&= & -{1\over ay}\ 
{}_2F_1\bigg({a,\alpha_1 \atop \beta_1} \bigg|x\bigg) +{1\over ay(1-y)^{a}}\ {}_2F_1\bigg({a,\alpha_1 \atop \beta_1} \bigg|{x\over 1-y}\bigg)
\end{eqnarray}
where $a \neq 0; \, \alpha_1 \in {\Bbb C}; \, \beta_1 \in {\Bbb C} \backslash {\Bbb Z}_0^-$ and
\begin{eqnarray}\label{eq6}
F_2(1;\alpha_1,1;\beta_1,2;x,y)&= & \,
{\alpha_1 \over \beta_1y}\bigg[{x\over 1-y}\ 
{}_3F_2\bigg( {\alpha_1+1,1,1 \atop \beta_1+1,2}\bigg|{x\over 1-y}\bigg)- x\ {}_3F_2\bigg({\alpha_1+1,1,1 \atop \beta_1+1,2}\bigg|x\bigg)\bigg]-{\ln(1-y)\over y}\nonumber\\
\end{eqnarray}
where $\alpha_1 \in {\Bbb C}; \, \beta_1 \in {\Bbb C} \backslash {\Bbb Z}_0^-.$}
\noindent  The present work is devoted to compute $F_2(\sigma;\alpha_1,1;\beta_1,2;x,y)$ explicitly for different values $\sigma,\alpha_1,\beta_1$ of the function parameters using (\ref{eq5}) and (\ref{eq6}). This is mostly done using many of the reduction formulas of $_2F_1$ and $_3F_2$ listed in \cite{Prudnikov} and other sources of special functions (\cite{Exton}, \cite{erdely1}-\cite{erdely2}, \cite{slater}, \cite{sri}), we begun here to tabulate reduction and transformation formulas for $F_2$. First, in Tables I and II, we give the corrections to some formulas misprinted in the classical monograph by Prudnikov et al. \cite{Prudnikov}.
\begin{table}[h1]
\begin{center}
\caption{Correction to some formulas for ${}_2F_1$ reported in the classical work of Prudnikov et al \cite{Prudnikov}. }
\vspace{0.1in}
\begin{tabular*}{1.0\textwidth}{@{\extracolsep{\fill}} | l || l || l || l | }
  \hline
\textbf{$\sigma$} &  $\alpha$    & $\beta$  & $_2F_1(\sigma,\alpha;\beta;z)$ \\
  \hline\hline
$\frac{5}{2}$ & $4$ & $1$ & $\frac{1}{16}(16+72z+18z^2-z^3)(1-z)^{-\frac{11}{2}}$ \\
\hline
 \hline
$\frac{4}{5}$ & $1$ & $\frac{14}{5}$ & $\frac{9}{5x^5}-\frac{9}{25x^9}(1-x^5)\big[\ln(1-x^5)-5\ln(1-x)-\sqrt{5}\ln\frac{1-2^{-1}(\sqrt{5}-1)x+x^2}{1+2^{-1}(\sqrt{5}+1)x+x^2}-$\\
~&~&~&$\quad 2(10+2\sqrt{5})^\frac{1}{2}\arctan\frac{(10+2\sqrt{5})^\frac{1}{2}x}{4-(\sqrt{5}-1)x}-2(10-2\sqrt{5})^\frac{1}{2}\arctan\frac{(10-2\sqrt{5})^\frac{1}{2}x}{4+(\sqrt{5}+1)x}\big]\qquad\qquad\qquad [x=z^\frac{1}{5}]$ \\
\hline
 \hline
$\frac{5}{6}$ & $1$ & $\frac{17}{5}$ & $\frac{11}{6x^6}+\frac{55}{36x^{11}}(1-x^6)(\ln\frac{1-x}{1+x}+\frac{1}{2}\ln\frac{1-x+x^2}{1+x+x^2}+\sqrt{3}\arctan\frac{3^\frac{1}{2}x}{1-x^2}\big]$ \\
\hline
 \hline
$1$ & $\frac{7}{2}$ & $\frac{9}{2}$ & -$\frac{7}{15z^3}\big[15+15z+3z^2-\frac{15\tanh^{-1}\sqrt{z}}{\sqrt{z}}\big]$ \\
\hline
 \hline
$1$ & $b$ & $b-m$ & $\frac{b-m-1}{b-1}\sum_{k=0}^{m-1}\frac{(-m)_k}{(2-b)_k}(1-z)^{-k-1}-\frac{m!}{(1-b)_m}(z-1)^{-m-1}$ \\
\hline
 \hline
$-\frac{n}{2}$ & $\frac{1-n}{2}$ & $1-n$ & $2^{-n}(1+\sqrt{1-z})^n,\qquad\qquad(n\neq{1, 2})$ \\
\hline
 \hline
\end{tabular*}
\end{center}
\end{table}
\begin{table}[h2]
\begin{center}
\caption{Correction to some formulas for $_3F_2$ reported in the classical work of Prudnikov et al \cite{Prudnikov}. }
\vspace{0.1in}
\begin{tabular*}{1.0\textwidth}{@{\extracolsep{\fill}} | l | l | l || l| l|| l | }
  \hline
\textbf{$a_1$} &  $a_2$    & $a_3$  & $b_1$& $b_2$  & $_3F_2(a_1,a_2,a_3;b_1,b_2;z)$ \\
  \hline\hline
$\frac{1}{4}$ &  $1$    & $1$  & $\frac{5}{4}$& $2$  & $\frac{1}{3z}\big[\ln(1-z)+z^\frac{3}{4}\big(\ln\frac{1+z^\frac{1}{4}}{1-z^\frac{1}{4}}+2\arctan{z^\frac{1}{4}}\big)\big]$\\
\hline
  \hline
\end{tabular*}
\end{center}
\end{table}
\section{Special values of the Appell Hypergeometric Functions $F_2$}

\noindent In the next, we tabulate the explicit computations of  $F_2(\sigma;\alpha_1,1;\beta_1,2;x,y)$ for different values of the function parameters $\sigma,\alpha_1$, and $\beta_1$. Since the role of $\alpha_1, \beta_1,x$ and $\alpha_2, \beta_2,y$ in $F_2(\sigma;\alpha_1,\alpha_2;\beta_1,\beta_2;x,y)$ are interchanged, similar tables can be obtain for  $F_2(\sigma;1,\alpha_2;2,\beta_2;x,y)$. Even-though the table is given only for $\alpha_2=1$ and $\beta_2=2$, the table can be used for a wider variety of cases. This can be notice from the following 
properties.
\begin{enumerate}
  \item $F_2(\sigma,\alpha_1,\alpha_2;\beta_1,\beta_2;x,y)=F_2(\sigma,\alpha_2,\alpha_1;\beta_2,\beta_1;y,x)$.
  \item $F_2(\sigma,\alpha_1,\alpha_2;\beta_1,\beta_2;x,y)=(1-x)^{-\sigma}F_2(\sigma,\beta_1-\alpha_1,\alpha_2;\beta_1,\beta_2;{x\over x-1},{y\over 1-x})=(1-x-y)^{-\sigma}F_2(\sigma,\beta_1-\alpha_1,\beta_2-\alpha_2;\beta_1,\beta_2;{x\over x+y-1},{y\over x+y-1})$.
  \item $F_1(\alpha;\beta,\beta',\gamma;x,y)=\left({x\over y}\right)^{\beta'}F_2(\beta+\beta';\alpha,\beta';\gamma,\beta+\beta';x,1-{x\over y})=\left({y\over x}\right)^{\beta}F_2(\beta+\beta';\alpha,\beta;\gamma,\beta+\beta';y,1-{y\over x}).$
 \end{enumerate}  
\noindent Further, the analytic expressions presented in the following tables can be useful in many applications of mathematical physics including the computation of the generalized Hubbell rectangular source integrals, elliptic integrals, and the radiation fields (\cite{beckert}, \cite{taraso1} to \cite{opps}, \cite{salman} to \cite{trasov2}).  

\begin{center}
\begin{tabular*}{1.0\textwidth}{@{\extracolsep{\fill}} | l || l | l || l| l|| l | }
  \hline
\textbf{$\sigma$} &  $\alpha_1$    & $\alpha_2$  & $\beta_1$& $\beta_2$  & $F_2(\sigma;\alpha_1,\alpha_2;\beta_1,\beta_2;x,y)$ \\
  \hline\hline
$\frac{1}{2}$ &  -$\frac{1}{2}$    & $1$  & -$\frac{1}{2}$& $2$  & $\frac{2}{y}[(1-x)^\frac{1}{2}-(1-x-y)^\frac{1}{2}]$ \\
\hline
  \hline
$\frac{1}{2}$ &  -$\frac{1}{2}$    & $1$  & $\frac{1}{2}$& $2$  & $\frac{2}{y}\big[\sqrt{1-x}-\sqrt{1-x-y}+\sqrt{x}(\arcsin\sqrt{x}-\arcsin\sqrt{\frac{x}{1-y}})\big]$\\
\hline
  \hline
$\frac{1}{2}$ &  $\frac{1}{2}$    & $1$  & -$\frac{1}{2}$& $2$  & $\frac{2}{y}[(1-x)^{-\frac{1}{2}}-(1-y)(1-x-y)^{-\frac{1}{2}}]$ \\
\hline
  \hline
$\frac{1}{2}$ &  $\frac{1}{2}$    & $1$  & $\frac{3}{2}$& $2$  & $\frac{1}{y}\big[\frac{\arcsin\sqrt{x}-(1-y)\arcsin\sqrt{\frac{x}{1-y}}}{\sqrt{x}}+\sqrt{1-x}-\sqrt{1-x-y}\big]$\\
\hline
  \hline
$\frac{1}{2}$ &  $\frac{1}{2}$    & $1$  & $\frac{5}{2}$& $2$  & $\frac{3}{8xy}\big[(1+2x)\sqrt{1-x}-(1-y+2x)\sqrt{1-x-y}-\frac{(1-4x)\arcsin\sqrt{x}-(1-y)(1-y-4x)\arcsin\sqrt{\frac{x}{1-y}}}{\sqrt{x}}\big]$\\
\hline
  \hline
$\frac{1}{2}$ &  $\frac{1}{2}$    & $1$  & $\frac{7}{2}$& $2$  & $\frac{5}{64x^2y}\big[3\big(\frac{(1-4x+8x^2)\arcsin\sqrt{x}-(1-y)\big((1-y)^2-4x(1-y)+8x^2\big)}{\sqrt{x}}\big)-(3-10x-8x^2)\sqrt{1-x}$\\
~&~&~&~&~&$\quad +\big(3(1-y)^2-10x(1-y)-8x^2\big)\sqrt{1-x-y}\big]$\\
\hline
  \hline
\end{tabular*}
\end{center}
\begin{center}
\begin{tabular*}{1.0\textwidth}{@{\extracolsep{\fill}} | l || l | l || l| l|| l | }
  \hline
\textbf{$\sigma$} &  $\alpha_1$    & $\alpha_2$  & $\beta_1$& $\beta_2$  & $F_2(\sigma;\alpha_1,\alpha_2;\beta_1,\beta_2;x,y)$ \\
  \hline\hline
$\frac{1}{2}$ &  $3$    & $1$  & $\frac{1}{2}$& $2$  & $\frac{1}{4y}\big[15\sqrt{x}(\tanh^{-1}\sqrt{\frac{x}{1-y}}-\tanh^{-1}\sqrt{x})-\frac{\sqrt{1-y}\big(8(1-y)^2-25x(1-y)+15x^2\big)}{(1-x-y)^2}+\frac{8-25x+15x^2}{(1-x)^2}\big]$\\
\hline
  \hline
$\frac{1}{2}$ &  $3$    & $1$  & $1$& $2$  & $\frac{1}{4y}\big[\frac{8-24x+15x^2}{(1-x)^\frac{3}{2}}-\frac{8(1-y)^2-24x(1-y)+15x^2}{(1-y-x)^\frac{3}{2}}\big]$ \\
\hline
  \hline
$\frac{1}{2}$ &  $3$    & $1$  & $\frac{3}{2}$& $2$  & -$\frac{1}{8y}\big[\frac{\sqrt{1-y}\big(13(1-y)-15x\big)}{1-x-y}-\frac{13-15x}{1-x}+3\big(\frac{(1-y-5x)\tanh^{-1}\sqrt{\frac{x}{1-y}}-(1-5x)\tanh^{-1}\sqrt{x}}{\sqrt{x}}\big)\big]$\\
\hline
  \hline
$\frac{1}{2}$ &  $3$    & $1$  & $2$& $2$  & $\frac{1}{2y}\big[\frac{4-5x}{\sqrt{1-x}}-\frac{4-4y-5x}{\sqrt{1-y-x}}\big]$ \\
\hline
  \hline
$\frac{1}{2}$ &  $3$    & $1$  & $\frac{5}{2}$& $2$  & $\frac{3}{32xy}\big[\sqrt{1-y}(1-y-15x)-1+15x-\frac{\big((1-y)^2+6x(1-y)-15x^2\big)\tanh^{-1}\sqrt{\frac{x}{1-y}}-(1+6x-15x^2)\tanh^{-1}\sqrt{x}}{\sqrt{x}}\big]$\\
\hline
  \hline
$\frac{1}{2}$ &  $3$    & $1$  & $\frac{7}{2}$& $2$  & $\frac{5}{64x^2y}\big[\sqrt{1-y}\big(3(1-y)^2+4x(1-y)-15x^2\big)-3-4x+15x^2$\\
~&~&~&~&~&$\quad -3\big(\frac{\big((1-y)^2+2x(1-y)+5x^2\big)(1-x-y)\tanh^{-1}\sqrt{\frac{x}{1-y}}-(1+2x+5x^2)(1-x)\tanh^{-1}\sqrt{x}}{\sqrt{x}}\big)\big]$\\
\hline
  \hline
\end{tabular*}
\end{center}
\begin{center}
\begin{tabular*}{1.0\textwidth}{@{\extracolsep{\fill}} | l || l | l || l| l|| l | }
  \hline
\textbf{$\sigma$} &  $\alpha_1$    & $\alpha_2$  & $\beta_1$& $\beta_2$  & $F_2(\sigma;\alpha_1,\alpha_2;\beta_1,\beta_2;x,y)$ \\
  \hline\hline
$1$ &  -$\frac{1}{2}$    & $1$  & $\frac{1}{2}$& $2$  & -$\frac{1}{y}\big[2\big(\sqrt{\frac{x}{1-y}}\tanh^{-1}(\sqrt{\frac{x}{1-y}})-\sqrt{x}\tanh^{-1}(\sqrt{x})\big)+\ln\big(\frac{1-y-x}{1-x}\big)\big]$ \\
\hline
\hline
$1$ &  -$\frac{1}{2}$    & $1$  & $1$& $2$  & $\frac{2}{y}\big[\sqrt{\frac{1-x-y}{1-y}}-\sqrt{1-x}+\ln\big(\frac{1+\sqrt{1-x}}{\sqrt{1-y}+\sqrt{1-x-y}}\big)\big]$\\
\hline
  \hline
$1$ &  -$\frac{1}{2}$    & $1$  & $\frac{3}{2}$& $2$  & -$\frac{1}{y}\big[\frac{1}{\sqrt{x}}\big(\frac{(1-y+x)\tanh^{-1}\sqrt{\frac{x}{1-y}}}{\sqrt{1-y}}-(1+x)\tanh^{-1}\sqrt{x}\big)-\ln\big(\frac{1-x}{1-x-y}\big)\big]$\\
\hline
  \hline
$1$ &  -$\frac{1}{2}$    & $1$  & $2$& $2$  & $\frac{2}{3xy}\big[y+\frac{(1-y+2x)\sqrt{1-x-y}}{\sqrt{1-y}}-(1+2x)\sqrt{1-x}+3x\ln\big(\frac{1+\sqrt{1-x}}{\sqrt{1-y}+\sqrt{1-x-y}}\big)\big]$\\
\hline
  \hline
$1$ &  -$\frac{1}{3}$    & $1$  & $\frac{2}{3}$& $2$  & $\frac{1}{y}\ln\big(\frac{1-x}{1-x-y}\big)-\frac{x^\frac{1}{3}}{2y}\big[(1-y)^{-\frac{1}{3}}\ln\big(\frac{1-x-y}{\big((1-y)^\frac{1}{3}-x^\frac{1}{3}\big)^3}\big)-\ln\big(\frac{1-x}{(1-x^\frac{1}{3})^3}\big)$\\
~&~&~&~&~&$\quad -2\sqrt{3}\big((1-y)^{-\frac{1}{3}}\arctan(\frac{\sqrt{3}x^\frac{1}{3}}{2(1-y)^\frac{1}{3}+x^\frac{1}{3}})-\arctan(\frac{\sqrt{3}x^\frac{1}{3}}{2+x^\frac{1}{3}})\big)\big]$\\
\hline
  \hline
$1$ &  -$\frac{1}{4}$    & $1$  & $\frac{3}{4}$& $2$  & -$\frac{1}{y}\big[\ln\big(\frac{1-x-y}{1-x}\big)+x^\frac{1}{4}\big((1-y)^{-\frac{1}{4}}\ln\big(\frac{(1-y)^\frac{1}{4}+x^\frac{1}{4}}{(1-y)^\frac{1}{4}-x^\frac{1}{4}}\big)-\ln\big(\frac{1+x^\frac{1}{4}}{1-x^\frac{1}{4}}\big)-\frac{2\arctan\big(\frac{x}{1-y}\big)^\frac{1}{4}}{(1-y)^\frac{1}{4}}+2\arctan(x^\frac{1}{4})\big)\big]$\\
\hline
  \hline
\end{tabular*}
\end{center}
\begin{center}
\begin{tabular*}{1.0\textwidth}{@{\extracolsep{\fill}} | l || l | l || l| l|| l | }
  \hline
\textbf{$\sigma$} &  $\alpha_1$    & $\alpha_2$  & $\beta_1$& $\beta_2$  & $F_2(\sigma;\alpha_1,\alpha_2;\beta_1,\beta_2;x,y)$ \\
  \hline\hline
$\frac{7}{6}$ &  $1$    & $1$  & $\frac{7}{6}$& $2$  & $\frac{1}{2x^\frac{1}{6}y}\big[\ln\frac{\big((1-y)^\frac{1}{6}+x^\frac{1}{6}\big)^2(1-x^\frac{1}{6})^2\big((1-y)^\frac{1}{3}+\big(x(1-y)\big)^\frac{1}{6}+x^\frac{1}{3}\big)(1-x^\frac{1}{6}+x^\frac{1}{3})}{\big((1-y)^\frac{1}{6}-x^\frac{1}{6}\big)^2(1+x^\frac{1}{6})^2\big((1-y)^\frac{1}{3}-\big(x(1-y)\big)^\frac{1}{6}+x^\frac{1}{3}\big)(1+x^\frac{1}{6}+x^\frac{1}{3})}$\\
~&~&~&~&~&$\quad +2\sqrt{3}\big(\arctan\frac{\sqrt{3}\big(x(1-y)\big)^\frac{1}{6}}{(1-y)^\frac{1}{3}-x^\frac{1}{3}}-\arctan\frac{\sqrt{3}x^\frac{1}{6}}{1-x^\frac{1}{3}}\big)\big]$\\
\hline
  \hline
$\frac{7}{6}$ &  $1$    & $1$  & $\frac{13}{6}$& $2$  & $\frac{7\big((1-y)^\frac{5}{6}-1\big)}{xy}-\frac{7}{6x^\frac{7}{6}y}\big[(1-x-y)\ln\frac{\big((1-y)^\frac{1}{6}+x^\frac{1}{6}\big)\sqrt{(1-y)^\frac{1}{3}+\big(x(1-y)\big)^\frac{1}{6}+x^\frac{1}{3}}}{\big((1-y)^\frac{1}{6}-x^\frac{1}{6}\big)\sqrt{(1-y)^\frac{1}{3}-\big(x(1-y)\big)^\frac{1}{6}+x^\frac{1}{3}}}$\\
~&~&~&~&~&$\quad -(1-x)\ln\frac{(1+x^\frac{1}{6})\sqrt{1+x^\frac{1}{6}+x^\frac{1}{3}}}{(1-x^\frac{1}{6})\sqrt{1-x^\frac{1}{6}+x^\frac{1}{3}}}+\sqrt{3}\big((1-x-y)\arctan\frac{\sqrt{3}\big(x(1-y)\big)^\frac{1}{6}}{(1-y)^\frac{1}{3}-x^\frac{1}{3}}-(1-x)\arctan\frac{\sqrt{3}x^\frac{1}{6}}{1-x^\frac{1}{3}}\big)\big]$\\
\hline
  \hline
\end{tabular*}
\end{center}
\begin{center}
\begin{tabular*}{1.0\textwidth}{@{\extracolsep{\fill}} | l || l | l || l| l|| l | }
  \hline
\textbf{$\sigma$} &  $\alpha_1$    & $\alpha_2$  & $\beta_1$& $\beta_2$  & $F_2(\sigma;\alpha_1,\alpha_2;\beta_1,\beta_2;x,y)$ \\
  \hline\hline
$\frac{7}{5}$ & $1$ & $1$ & $\frac{12}{5}$& $2$ & $\frac{7\big((1-y)^\frac{3}{5}-1\big)}{2xy}-\frac{7}{20x^\frac{7}{5}y}\big[(1-x-y)\ln\frac{1-x-y}{\big((1-y)^\frac{1}{5}-x^\frac{1}{5}\big)^5}-(1-x)\ln\frac{1-x}{(1-x^\frac{1}{5})^5}$\\
~&~&~&~&~&$\quad -\sqrt{5}\big((1-x-y)\ln\frac{(1-y)^\frac{2}{5}+\frac{\sqrt{5}+1}{2}\big(x(1-y)\big)^\frac{1}{5}+x^\frac{2}{5}}{(1-y)^\frac{2}{5}-\frac{\sqrt{5}-1}{2}\big(x(1-y)\big)^\frac{1}{5}+x^\frac{2}{5}}-(1-x)\ln\frac{1+\frac{\sqrt{5}+1}{2}x^\frac{1}{5}+x^\frac{2}{5}}{1-\frac{\sqrt{5}-1}{2}x^\frac{1}{5}+x^\frac{2}{5}}\big)$\\
~&~&~&~&~&$\quad\quad +2\sqrt{10-2\sqrt{5}}\big((1-x-y)\arctan\frac{\sqrt{10+2\sqrt{5}}x^\frac{1}{5}}{4(1-y)^\frac{1}{5}-(\sqrt{5}-1)x^\frac{1}{5}}-(1-x)\arctan\frac{\sqrt{10+2\sqrt{5}}x^\frac{1}{5}}{4-(\sqrt{5}-1)x^\frac{1}{5}}\big)$\\
~&~&~&~&~&$\quad\quad\quad -2\sqrt{10+2\sqrt{5}}\big((1-x-y)\arctan\frac{\sqrt{10-2\sqrt{5}}x^\frac{1}{5}}{4(1-y)^\frac{1}{5}+(\sqrt{5}+1)x^\frac{1}{5}}-(1-x)\arctan\frac{\sqrt{10-2\sqrt{5}}x^\frac{1}{5}}{4+(\sqrt{5}+1)x^\frac{1}{5}}\big)$\\
\hline
  \hline
$\frac{3}{2}$ &  -$\frac{1}{2}$    & $1$  & $\frac{1}{2}$& $2$  & $\frac{2}{y}[(1-y)^{-1}(1-x-y)^\frac{1}{2}-(1-x)^\frac{1}{2}]$ \\
\hline
  \hline
$\frac{3}{2}$ &  -$\frac{1}{2}$    & $1$  & $\frac{3}{2}$& $2$  & $\frac{1}{y}\big[\frac{\arcsin\sqrt{\frac{x}{1-y}}-\arcsin\sqrt{x}}{\sqrt{x}}+\frac{\sqrt{1-x-y}}{1-y}-\sqrt{1-x}\big]$\\
\hline
  \hline
$\frac{3}{2}$ &  -$\frac{1}{2}$    & $1$  & $\frac{5}{2}$& $2$  & $\frac{3}{8xy}\big[\frac{(1-y+2x)\sqrt{1-x-y}}{1-y}-(1+2x)\sqrt{1-x}-\frac{(1-y-4x)\arcsin\sqrt{\frac{x}{1-y}}-(1-4x)\arcsin\sqrt{x}}{\sqrt{x}}\big]$\\
\hline
  \hline
\end{tabular*}
\end{center}
\begin{center}
\begin{tabular*}{1.0\textwidth}{@{\extracolsep{\fill}} | l || l | l || l| l|| l | }
  \hline
\textbf{$\sigma$} &  $\alpha_1$    & $\alpha_2$  & $\beta_1$& $\beta_2$  & $F_2(\sigma;\alpha_1,\alpha_2;\beta_1,\beta_2;x,y)$ \\
  \hline\hline
$\frac{3}{2}$ & $2$ & $1$ & $\frac{7}{2}$& $2$ & $\frac{15}{8x^2y}\big[\sqrt{1-y}\big(3(1-y)-x\big)-(3-x)-\frac{(1-x-y)\big(3(1-y)+x\big)\tanh^{-1}\sqrt{\frac{x}{1-y}}-(1-x)(3+x)\tanh^{-1}\sqrt{x}}{\sqrt{x}}\big]$\\
\hline
  \hline
$\frac{3}{2}$ &  $2$    & $1$  & $4$& $2$  & $\frac{16}{5x^3y}\big[\big(4(1-y)+x\big)(1-x-y)^{\frac{3}{2}}-\big(4(1-y)-5x\big)(1-y)^{\frac{3}{2}}-(4+x)(1-x)^{\frac{3}{2}}+4-5x\big]$ \\
\hline
  \hline
$\frac{3}{2}$ &  $\frac{5}{2}$    & $1$  & -$\frac{1}{2}$& $2$  & $\frac{2}{y}\big[\frac{(1-y-6x)(1-y)^2}{(1-x-y)^{\frac{7}{2}}}-\frac{1-6x}{(1-x)^{\frac{7}{2}}}\big]$ \\
\hline
  \hline
$\frac{3}{2}$ &  $\frac{5}{2}$    & $1$  & $\frac{1}{2}$& $2$  & $\frac{2}{y}[(1-y)^2(1-x-y)^{-\frac{5}{2}}-(1-x)^{-\frac{5}{2}}]$ \\
\hline
  \hline
$\frac{3}{2}$ &  $\frac{5}{2}$    & $1$  & $\frac{3}{2}$& $2$  & $\frac{2}{3y}\big[\frac{3(1-y)-2x}{(1-x-y)^{\frac{3}{2}}}-\frac{3-2x}{(1-x)^{\frac{3}{2}}}\big]$ \\
\hline
  \hline
$\frac{3}{2}$ & $\frac{5}{2}$ & $1$ & $\frac{7}{2}$& $2$ & $\frac{5}{4x^2y}\big[3\frac{(1-y)^2\arcsin\sqrt{\frac{x}{1-y}}-\arcsin\sqrt{x}}{\sqrt{x}}-\big(3(1-y)+2x\big)\sqrt{1-x-y}+(3+2x)\sqrt{1-x}\big]$\\
\hline
  \hline
$\frac{3}{2}$ &  $3$    & $1$  & -$\frac{1}{2}$& $2$  & $\frac{2}{y}\big[\frac{(1-y-7x)(1-y)^{\frac{5}{2}}}{(1-x-y)^{4}}-\frac{1-7x}{(1-x)^{4}}\big]$ \\
\hline
  \hline
\end{tabular*}
\end{center}
\begin{center}
\begin{tabular*}{1.0\textwidth}{@{\extracolsep{\fill}} | l || l | l || l| l|| l | }
  \hline
\textbf{$\sigma$} &  $\alpha_1$    & $\alpha_2$  & $\beta_1$& $\beta_2$  & $F_2(\sigma;\alpha_1,\alpha_2;\beta_1,\beta_2;x,y)$ \\
  \hline\hline
$\frac{5}{3}$ & $1$ & $1$ & $\frac{5}{3}$& $2$ & $\frac{1}{2x^\frac{2}{3}y}\big[\ln\frac{(1-x-y)(1-x^\frac{1}{3})^3}{(1-x)\big((1-y)^\frac{1}{3}-x^\frac{1}{3}\big)^3}-2\sqrt{3}\big(\arctan\frac{\sqrt{3}x^\frac{1}{3}}{2(1-y)^\frac{1}{3}+x^\frac{1}{3}}-\arctan\frac{\sqrt{3}x^\frac{1}{3}}{2+x^\frac{1}{3}}\big)\big]$\\
\hline
  \hline
$\frac{5}{3}$ & $1$ & $1$ & $\frac{8}{3}$& $2$ & $\frac{5\big((1-y)^\frac{1}{3}-1\big)}{2xy}-\frac{5}{6x^\frac{5}{3}y}\big[(1-x-y)\ln\frac{1-x-y}{\big((1-y)^\frac{1}{3}-x^\frac{1}{3}\big)^3}-(1-x)\ln\frac{1-x}{(1-x^\frac{1}{3})^3}$\\
~&~&~&~&~&$\quad -2\sqrt{3}\big((1-x-y)\arctan\frac{\sqrt{3}x^\frac{1}{3}}{2(1-y)^\frac{1}{3}+x^\frac{1}{3}}-(1-x)\arctan\frac{\sqrt{3}x^\frac{1}{3}}{2+x^\frac{1}{3}}\big)\big]$\\
\hline
  \hline
$\frac{7}{4}$ &  $\frac{1}{4}$    & $1$  & $\frac{1}{2}$& $2$  & $\frac{2\sqrt{2}}{3y}\big[\frac{\sqrt{\sqrt{1-y}+\sqrt{1-x-y}}}{\sqrt{(1-x-y)(1-y)}}-\frac{\sqrt{1+\sqrt{1-x}}}{\sqrt{1-x}}\big]$\\
\hline
  \hline
$\frac{7}{4}$ &  $\frac{1}{4}$    & $1$  & $\frac{3}{2}$& $2$  & $\frac{4}{3\sqrt{x}y}\big[\frac{\sqrt{\sqrt{1-y}+\sqrt{x}}-\sqrt{\sqrt{1-y}-\sqrt{x}}}{\sqrt{1-y}}-\sqrt{1+\sqrt{x}}+\sqrt{1-\sqrt{x}}\big]$\\
\hline
  \hline
$\frac{7}{4}$ & $1$ & $1$ & $\frac{7}{4}$& $2$ & $\frac{1}{x^\frac{3}{4}y}\big[\ln\frac{\big((1-y)^\frac{1}{4}+x^\frac{1}{4}\big)(1-x^\frac{1}{4})}{\big((1-y)^\frac{1}{4}-x^\frac{1}{4}\big)(1+x^\frac{1}{4})}-2\big(\arctan(\frac{x}{1-y})^\frac{1}{4}-\arctan{x^\frac{1}{4}}\big)\big]$\\
\hline
  \hline
$\frac{7}{4}$ & $1$ & $1$ & $\frac{11}{4}$& $2$ & $\frac{7\big((1-y)^\frac{1}{4}-1\big)}{3xy}-\frac{7}{4x^\frac{7}{4}y}\big[(1-x-y)\ln\frac{(1-y)^\frac{1}{4}+x^\frac{1}{4}}{(1-y)^\frac{1}{4}-x^\frac{1}{4}}-(1-x)\ln\frac{1+x^\frac{1}{4}}{1-x^\frac{1}{4}}$\\
\hline
  \hline
\end{tabular*}
\end{center}
\begin{center}
\begin{tabular*}{1.0\textwidth}{@{\extracolsep{\fill}} | l || l | l || l| l|| l | }
  \hline
\textbf{$\sigma$} &  $\alpha_1$    & $\alpha_2$  & $\beta_1$& $\beta_2$  & $F_2(\sigma;\alpha_1,\alpha_2;\beta_1,\beta_2;x,y)$ \\
  \hline\hline
$2$ &  -$\frac{1}{2}$    & $1$  & $3$& $2$  & $\frac{8}{15x^2y}\big[\frac{(1-x-y)^\frac{5}{2}}{(1-y)^\frac{3}{2}}-(1-x)^\frac{5}{2}+y\big]$\\
\hline
  \hline
$2$ &  -$\frac{1}{2}$    & $1$  & $\frac{7}{2}$& $2$  & $\frac{5}{48x^2y}\big[\frac{3}{\sqrt{x}}\big(\frac{(1-x-y)^3\tanh^{-1}\sqrt{\frac{x}{1-y}}}{(1-y)^\frac{3}{2}}-(1-x)^3\tanh^{-1}\sqrt{x}\big)-\frac{3(1-y)^2-8x(1-y)-3x^2}{1-y}+3-8x-3x^2\big]$\\
\hline
  \hline
$2$ &  -$\frac{1}{2}$    & $1$  & $4$& $2$  & $\frac{8}{35x^3y}\big[2\big((1-y)^2-1\big)+7xy-2\big(\frac{(1-x-y)^\frac{7}{2}}{(1-y)^\frac{3}{2}}-(1-x)^\frac{7}{2}\big)\big]$\\
\hline
  \hline
$2$ &  $\frac{1}{8}$    & $1$  & $\frac{9}{8}$& $2$  & $\frac{1}{8x^\frac{1}{8}y}\big[(1-y)^{-\frac{7}{8}}\ln\frac{(1-y)^\frac{1}{8}+x^\frac{1}{8}}{(1-y)^\frac{1}{8}-x^\frac{1}{8}}-\ln\frac{1+x^\frac{1}{8}}{1-x^\frac{1}{8}}+2\big((1-y)^{-\frac{7}{8}}\arctan\big(\frac{x}{1-y}\big)^\frac{1}{8}-\arctan(x^\frac{1}{8})\big)$\\
~&~&~&~&~&$\quad +\sqrt{2}\big((1-y)^{-\frac{7}{8}}\arctan\frac{\sqrt{2}\big(x(1-y)\big)^\frac{1}{8}}{(1-y)^\frac{1}{4}-x^\frac{1}{4}}-\arctan\frac{\sqrt{2}x^\frac{1}{8}}{1-x^\frac{1}{4}}\big)$\\
~&~&~&~&~&$\quad\quad -2^{-\frac{1}{2}}\big((1-y)^{-\frac{7}{8}}\ln\frac{(1-y)^\frac{1}{4}-\sqrt{2}\big(x(1-y)\big)^\frac{1}{8}+x^\frac{1}{4}}{(1-y)^\frac{1}{4}+\sqrt{2}\big(x(1-y)\big)^\frac{1}{8}+x^\frac{1}{4}}-\ln\frac{1-\sqrt{2}x^\frac{1}{8}+x^\frac{1}{4}}{1+\sqrt{2}x^\frac{1}{8}+x^\frac{1}{4}}\big]$\\
\hline
  \hline
\end{tabular*}
\end{center}
\begin{center}
\begin{tabular*}{1.0\textwidth}{@{\extracolsep{\fill}} | l || l | l || l| l|| l | }
  \hline
\textbf{$\sigma$} &  $\alpha_1$    & $\alpha_2$  & $\beta_1$& $\beta_2$  & $F_2(\sigma;\alpha_1,\alpha_2;\beta_1,\beta_2;x,y)$ \\
  \hline\hline
$2$ & $\frac{3}{8}$ & $1$ & $\frac{11}{8}$& $2$ & $\frac{3}{8x^\frac{3}{8}y}\big[(1-y)^{-\frac{5}{8}}\ln\frac{(1-y)^\frac{1}{8}+x^\frac{1}{8}}{(1-y)^\frac{1}{8}-x^\frac{1}{8}}-\ln\frac{1+x^\frac{1}{8}}{1-x^\frac{1}{8}}$\\
~&~&~&~&~&$\quad -2^{-\frac{1}{2}}\big((1-y)^{-\frac{5}{8}}\ln\frac{(1-y)^\frac{1}{4}+\sqrt{2}\big(x(1-y)\big)^\frac{1}{8}+x^\frac{1}{4}}{(1-y)^\frac{1}{4}-\sqrt{2}\big(x(1-y)\big)^\frac{1}{8}+x^\frac{1}{4}}-\ln\frac{1+\sqrt{2}x^\frac{1}{8}+x^\frac{1}{4}}{1-\sqrt{2}x^\frac{1}{8}+x^\frac{1}{4}}\big)$\\
~&~&~&~&~&$\quad\quad +\sqrt{2}\big((1-y)^{-\frac{5}{8}}\arctan\frac{\sqrt{2}\big(x(1-y)\big)^\frac{1}{8}}{(1-y)^\frac{1}{4}-x^\frac{1}{4}}-\arctan\frac{\sqrt{2}x^\frac{1}{8}}{1-x^\frac{1}{4}}\big)-2\big((1-y)^{-\frac{5}{8}}\arctan(\frac{x}{1-y})^\frac{1}{8}-\arctan{x^\frac{1}{8}}\big)\big]$\\
\hline
  \hline
$2$ & $\frac{2}{5}$ & $1$ & $\frac{7}{5}$& $2$ & $\frac{1}{10x^\frac{2}{3}y}\big[(1-y)^{-\frac{3}{5}}\ln\frac{1-x-y}{\big((1-y)^\frac{1}{5}-x^\frac{1}{5}\big)^5}-\ln\frac{1-x}{(1-x^\frac{1}{5})^5}$\\
~&~&~&~&~&$\quad -\sqrt{5}\big((1-y)^{-\frac{3}{5}}\ln\frac{(1-y)^\frac{2}{5}+\frac{\sqrt{5}+1}{2}\big(x(1-y)\big)^\frac{1}{5}+x^\frac{2}{5}}{(1-y)^\frac{2}{5}-\frac{\sqrt{5}-1}{2}\big(x(1-y)\big)^\frac{1}{5}+x^\frac{2}{5}}-\ln\frac{1+\frac{\sqrt{5}+1}{2}x^\frac{1}{5}+x^\frac{2}{5}}{1-\frac{\sqrt{5}-1}{2}x^\frac{1}{5}+x^\frac{2}{5}}\big)$\\
~&~&~&~&~&$\quad\quad +2\sqrt{10-2\sqrt{5}}\big((1-y)^{-\frac{3}{5}}\arctan\frac{\sqrt{10+2\sqrt{5}}x^\frac{1}{5}}{4(1-y)^\frac{1}{5}-(\sqrt{5}-1)x^\frac{1}{5}}-\arctan\frac{\sqrt{10+2\sqrt{5}}x^\frac{1}{5}}{4-(\sqrt{5}-1)x^\frac{1}{5}}\big)$\\
\hline
  \hline
\end{tabular*}
\end{center}
\begin{center}
\begin{tabular*}{1.0\textwidth}{@{\extracolsep{\fill}} | l || l | l || l| l|| l | }
  \hline
\textbf{$\sigma$} &  $\alpha_1$    & $\alpha_2$  & $\beta_1$& $\beta_2$  & $F_2(\sigma;\alpha_1,\alpha_2;\beta_1,\beta_2;x,y)$ \\
  \hline\hline
$2$ & $\frac{3}{4}$ & $1$ & $\frac{7}{4}$& $2$ & $\frac{3}{4x^\frac{3}{4}y}\big[(1-y)^{-\frac{1}{4}}\ln\frac{(1-y)^\frac{1}{4}+x^\frac{1}{4}}{(1-y)^\frac{1}{4}-x^\frac{1}{4}}-\ln\frac{1+x^\frac{1}{4}}{1-x^\frac{1}{4}}-2\big((1-y)^{-\frac{1}{4}}\arctan(\frac{x}{1-y})^\frac{1}{4}-\arctan{x^\frac{1}{4}}\big)\big]$\\
\hline
  \hline
$2$ & $\frac{3}{4}$ & $1$ & $\frac{11}{4}$& $2$ & -$\frac{21}{16x^\frac{7}{4}y}\big[\frac{1-x-y}{(1-y)^\frac{1}{4}}\ln\frac{(1-y)^\frac{1}{4}+x^\frac{1}{4}}{(1-y)^\frac{1}{4}-x^\frac{1}{4}}-(1-x)\ln\frac{1+x^\frac{1}{4}}{1-x^\frac{1}{4}}-2\big(\frac{1-x-y}{(1-y)^\frac{1}{4}}\arctan(\frac{x}{1-y})^\frac{1}{4}-(1-x)\arctan{x^\frac{1}{4}}\big)\big]$\\
\hline
  \hline
$2$ & $\frac{4}{5}$ & $1$ & $\frac{9}{5}$& $2$ & $\frac{1}{5x^\frac{4}{5}y}\big[(1-y)^{-\frac{1}{5}}\ln\frac{1-x-y}{\big((1-y)^\frac{1}{5}-x^\frac{1}{5}\big)^5}-\ln\frac{1-x}{(1-x^\frac{1}{5})^5}$\\
~&~&~&~&~&$\quad -\sqrt{5}\big((1-y)^{-\frac{1}{5}}\ln\frac{(1-y)^\frac{2}{5}-\frac{\sqrt{5}-1}{2}(x(1-y))^\frac{1}{5}+x^\frac{2}{5}}{(1-y)^\frac{2}{5}+\frac{\sqrt{5}+1}{2}(x(1-y))^\frac{1}{5}+x^\frac{2}{5}}-\ln\frac{1-\frac{\sqrt{5}-1}{2}x^\frac{1}{5}+x^\frac{2}{5}}{1+\frac{\sqrt{5}+1}{2}x^\frac{1}{5}+x^\frac{2}{5}}\big)$\\
~&~&~&~&~&$\quad\quad -2\sqrt{10+2\sqrt{5}}\big((1-y)^{-\frac{1}{5}}\arctan\frac{\sqrt{10+2\sqrt{5}}x^\frac{1}{5}}{4(1-y)^\frac{1}{5}-(\sqrt{5}-1)x^\frac{1}{5}}-\arctan\frac{\sqrt{10+2\sqrt{5}}x^\frac{1}{5}}{4-(\sqrt{5}-1)x^\frac{1}{5}}\big)$\\
~&~&~&~&~&$\quad\quad\quad -2\sqrt{10-2\sqrt{5}}\big((1-y)^{-\frac{1}{5}}\arctan\frac{\sqrt{10-2\sqrt{5}}x^\frac{1}{5}}{4(1-y)^\frac{1}{5}+(\sqrt{5}+1)x^\frac{1}{5}}-\arctan\frac{\sqrt{10-2\sqrt{5}}x^\frac{1}{5}}{4+(\sqrt{5}+1)x^\frac{1}{5}}\big)\big]$\\
\hline
  \hline
\end{tabular*}
\end{center}
\begin{center}
\begin{tabular*}{1.0\textwidth}{@{\extracolsep{\fill}} | l || l | l || l| l|| l | }
  \hline
\textbf{$\sigma$} &  $\alpha_1$    & $\alpha_2$  & $\beta_1$& $\beta_2$  & $F_2(\sigma;\alpha_1,\alpha_2;\beta_1,\beta_2;x,y)$ \\
  \hline\hline
$2$ & $\frac{3}{2}$ & $1$ & $\frac{5}{2}$& $2$ & $\frac{3}{x^\frac{3}{2}y}\big[\sqrt{1-y}\tanh^{-1}\sqrt{\frac{x}{1-y}}-\tanh^{-1}\sqrt{x}\big]$\\
\hline
  \hline
$2$ &  $\frac{3}{2}$    & $1$  & $3$& $2$  & $\frac{4}{x^2y}\big[\sqrt{1-x}-\sqrt{(1-y)(1-x-y)}-y\big]$ \\
\hline
  \hline
$2$ & $\frac{3}{2}$ & $1$ & $\frac{7}{2}$& $2$ & -$\frac{15}{2x^2y}\big[y+\frac{\sqrt{1-y}(1-x-y)\tanh^{-1}\sqrt{\frac{x}{1-y}}-(1-x)\tanh^{-1}\sqrt{x}}{\sqrt{x}}\big]$\\
\hline
  \hline
$2$ &  $\frac{3}{2}$    & $1$  & $4$& $2$  & $\frac{2}{x^3y}\big[8(1-y)^{\frac{1}{2}}\big((1-x-y)^{\frac{3}{2}}-(1-y)^{\frac{3}{2}}\big)-8\big((1-x)^{\frac{3}{2}}-1\big)-12xy\big]$ \\
\hline
  \hline
$2$ & $2$ & $1$ & -$\frac{1}{2}$& $2$ & $\frac{1}{2y}\big[\frac{2(1-y)^2-14x(1-y)-3x^2}{(1-x-y)^3}-\frac{2-14x-3x^2}{(1-x)^3}-15x^\frac{3}{2}\big(\frac{(1-y)\arcsin\sqrt{\frac{x}{1-y}}}{(1-x-y)^\frac{7}{2}}-\frac{\arcsin\sqrt{x}}{(1-x)^\frac{7}{2}}\big)\big]$\\
\hline
  \hline
$2$ & $2$ & $1$ & $\frac{1}{2}$& $2$ & $\frac{1}{2y}\big[\frac{2(1-y)+x}{(1-x-y)^2}-\frac{2+x}{(1-x)^2}+3\sqrt{x}\big(\frac{(1-y)\arcsin\sqrt{\frac{x}{1-y}}}{(1-x-y)^\frac{5}{2}}-\frac{\arcsin\sqrt{x}}{(1-x)^\frac{5}{2}}\big)\big]$\\
\hline
  \hline
$2$ &  $2$    & $1$  & $1$& $2$  & $\frac{1}{y}[(1-y)(1-x-y)^{-2}-(1-x)^{-2}]$ \\
\hline
  \hline
\end{tabular*}
\end{center}
\begin{center}
\begin{tabular*}{1.0\textwidth}{@{\extracolsep{\fill}} | l || l | l || l| l|| l | }
  \hline
\textbf{$\sigma$} &  $\alpha_1$    & $\alpha_2$  & $\beta_1$& $\beta_2$  & $F_2(\sigma;\alpha_1,\alpha_2;\beta_1,\beta_2;x,y)$ \\
  \hline\hline
$2$ &  $\frac{7}{2}$    & $1$  & $3$& $2$  & $\frac{8}{15x^2y}\big[\frac{(1-y)^\frac{5}{2}}{(1-x-y)^{\frac{3}{2}}}-\frac{1}{(1-x)^{\frac{3}{2}}}+y\big]$ \\
\hline
  \hline
$2$ &  $\frac{7}{2}$    & $1$  & $4$& $2$  & $\frac{8}{5x^3y}\big[\frac{2(1-y)^\frac{5}{2}}{\sqrt{1-x-y}}-\frac{2}{\sqrt{1-x}}-2\big((1-y)^2-1\big)+xy\big]$ \\
\hline
  \hline
$2$ & $\frac{7}{2}$ & $1$ & $\frac{9}{2}$& $2$ & $\frac{7}{3x^3y}\big[3\frac{(1-y)^\frac{5}{2}\tanh^{-1}\sqrt{\frac{x}{1-y}}-\tanh^{-1}\sqrt{x}}{\sqrt{x}}-3\big((1-y)^2-1\big)-xy\big]$\\
\hline
  \hline
$2$ & $4$ & $1$ & -$\frac{1}{2}$& $2$ & $\frac{1}{16y}\big[\frac{16(1-y)^4-208x(1-y)^3-165x^2(1-y)^2+50x^3(1-y)-8x^4}{(1-x-y)^5}-\frac{16-208x-165x^2+50x^3-8x^4}{(1-x)^5}$\\
~&~&~&~&~&$\quad -315x^\frac{3}{2}\big(\frac{(1-y)^3\arcsin\sqrt{\frac{x}{1-y}}}{(1-x-y)^\frac{11}{2}}-\frac{\arcsin\sqrt{x}}{(1-x)^\frac{11}{2}}\big)\big]$\\
\hline
  \hline
$2$ & $4$ & $1$ & $\frac{1}{2}$& $2$ & $\frac{1}{48y}\big[\frac{48(1-y)^3+87x(1-y)^2-38x^2(1-y)+8x^3}{(1-x-y)^4}-\frac{48+87x-38x^2+8x^3}{(1-x)^4}+105\sqrt{x}\big(\frac{(1-y)^3\arcsin\sqrt{\frac{x}{1-y}}}{(1-x-y)^\frac{9}{2}}-\frac{\arcsin\sqrt{x}}{(1-x)^\frac{9}{2}}\big)\big]$\\
\hline
  \hline
$2$ &  $4$    & $1$  & $1$& $2$  & $\frac{1}{y}[(1-y)^3(1-x-y)^{-4}-(1-x)^{-4}]$ \\
\hline
  \hline
\end{tabular*}
\end{center}
\begin{center}
\begin{tabular*}{1.0\textwidth}{@{\extracolsep{\fill}} | l || l | l || l| l|| l | }
  \hline
\textbf{$\sigma$} &  $\alpha_1$    & $\alpha_2$  & $\beta_1$& $\beta_2$  & $F_2(\sigma;\alpha_1,\alpha_2;\beta_1,\beta_2;x,y)$ \\
  \hline\hline
$\frac{5}{2}$ &  $\frac{3}{2}$    & $1$  & $\frac{1}{2}$& $2$  & $\frac{2}{3y}\big[\frac{1-y+2x}{(1-x-y)^\frac{5}{2}}-\frac{1+2x}{(1-x)^\frac{5}{2}}\big]$ \\
\hline
  \hline
$\frac{5}{2}$ &  $\frac{3}{2}$    & $1$  & $\frac{3}{2}$& $2$  & $\frac{2}{3y}[(1-x-y)^{-\frac{3}{2}}-(1-x)^{-\frac{3}{2}}]$ \\
\hline
  \hline
$\frac{5}{2}$ & $\frac{3}{2}$ & $1$ & $\frac{5}{2}$& $2$ & $\frac{2}{xy}\big[(1-x-y)^{-\frac{1}{2}}-(1-x)^{-\frac{1}{2}}-\frac{\arcsin\sqrt{\frac{x}{1-y}}-\arcsin\sqrt{x}}{\sqrt{x}}\big]$\\
\hline
  \hline
$\frac{5}{2}$ & $\frac{3}{2}$ & $1$ & $\frac{7}{2}$& $2$ & $\frac{5}{2x^2y}\big[\frac{\big(3(1-y)-2x\big)\arcsin\sqrt{\frac{x}{1-y}}-(3-2x)\arcsin\sqrt{x}}{\sqrt{x}}-3(\sqrt{1-x-y}-\sqrt{1-x})\big]$\\
\hline
  \hline
$\frac{5}{2}$ &  $2$    & $1$  & -$\frac{1}{2}$& $2$  & $\frac{2}{3y}\big[\frac{\sqrt{1-y}\big((1-y)^2-10x(1-y)-15x^2\big)}{(1-x-y)^4}-\frac{1-10x-15x^2}{(1-x)^4}\big]$ \\
\hline
  \hline
$\frac{5}{2}$ &  $2$    & $1$  & $\frac{1}{2}$& $2$  & $\frac{2}{3y}\big[\frac{\sqrt{1-y}(1-y+3x)}{(1-x-y)^3}-\frac{1+3x}{(1-x)^3}\big]$ \\
\hline
  \hline
$\frac{5}{2}$ &  $2$    & $1$  & $1$& $2$  & $\frac{1}{3y}\big[\frac{2(1-y)+x}{(1-x-y)^\frac{5}{2}}-\frac{2+x}{(1-x)^\frac{5}{2}}\big]$ \\
\hline
  \hline
\end{tabular*}
\end{center}
\begin{center}
\begin{tabular*}{1.0\textwidth}{@{\extracolsep{\fill}} | l || l | l || l| l|| l | }
  \hline
\textbf{$\sigma$} &  $\alpha_1$    & $\alpha_2$  & $\beta_1$& $\beta_2$  & $F_2(\sigma;\alpha_1,\alpha_2;\beta_1,\beta_2;x,y)$ \\
  \hline\hline
$\frac{5}{2}$ &  $4$    & $1$  & $3$& $2$  & $\frac{1}{3y}\big[\frac{2(1-y)-x}{(1-x-y)^\frac{5}{2}}-\frac{2-x}{(1-x)^\frac{5}{2}}\big]$ \\
\hline
  \hline
$\frac{5}{2}$ & $4$ & $1$ & $\frac{7}{2}$& $2$ & $\frac{5}{48x^2y}\big[\frac{\sqrt{1-y}\big(3(1-y)^2-2x(1-y)+3x^2\big)}{(1-x-y)^2}-\frac{3-2x+3x^2}{(1-x)^2}-3\big(\frac{(1+x-y)\tanh^{-1}\sqrt{\frac{x}{1-y}}-(1+x)\tanh^{-1}\sqrt{x}}{\sqrt{x}}\big)\big]$\\
\hline
  \hline
$3$ &  -$\frac{1}{2}$    & $1$  & $\frac{1}{2}$& $2$  & $\frac{1}{4y}\big[\frac{\big(2(1-y)-3x\big)}{(1-y)^2(1-x-y)}-\frac{2-3x}{1-x}-3\sqrt{x}(\frac{\tanh^{-1}\sqrt{\frac{x}{1-y}}}{(1-y)^\frac{5}{2}}-\tanh^{-1}\sqrt{x})\big]$\\
\hline
  \hline
$3$ &  -$\frac{1}{2}$    & $1$  & $1$& $2$  & $\frac{1}{4y}\big[\frac{2(1-y)-3x}{(1-y)^\frac{5}{2}\sqrt{1-x-y}}-\frac{2-3x}{\sqrt{1-x}}\big]$\\
\hline
  \hline
$3$ &  -$\frac{1}{2}$    & $1$  & $\frac{3}{2}$& $2$  & $\frac{1}{8y}\big[3\big((1-y)^{-2}-1\big)+\frac{1}{\sqrt{x}}\big(\frac{(1-y-3x)\tanh^{-1}\sqrt{\frac{x}{1-y}}}{(1-y)^\frac{5}{2}}-(1-3x)\tanh^{-1}\sqrt{x}\big)\big]$\\
\hline
  \hline
$3$ &  -$\frac{1}{2}$    & $1$  & $2$& $2$  & $\frac{1}{2y}[(1-y)^{-\frac{5}{2}}(1-x-y)^\frac{1}{2}-(1-x)^\frac{1}{2}]$ \\
\hline
  \hline
$3$ &  -$\frac{1}{2}$    & $1$  & $\frac{5}{2}$& $2$  & $\frac{3}{32xy}\big[\frac{1}{\sqrt{x}}\big(\frac{(1-y+3x)(1-y-x)\tanh^{-1}\sqrt{\frac{x}{1-y}}}{(1-y)^\frac{5}{2}}-(1+3x)(1-x)\tanh^{-1}\sqrt{x}\big)+1-3x-\frac{(1-y-3x)}{(1-y)^2}\big]$\\
\hline
  \hline
\end{tabular*}
\end{center}
\begin{center}
\begin{tabular*}{1.0\textwidth}{@{\extracolsep{\fill}} | l || l | l || l| l|| l | }
  \hline
\textbf{$\sigma$} &  $\alpha_1$    & $\alpha_2$  & $\beta_1$& $\beta_2$  & $F_2(\sigma;\alpha_1,\alpha_2;\beta_1,\beta_2;x,y)$ \\
  \hline\hline
$3$ & $\frac{3}{2}$ & $1$ & $3$& $2$ & $\frac{2}{x^2y}\big[\frac{2(1-y)-x}{\sqrt{(1-y)(1-x-y)}}-\frac{2-x}{\sqrt{1-x}}\big]$\\
\hline
  \hline
$3$ & $\frac{3}{2}$ & $1$ & $\frac{7}{2}$& $2$ & $\frac{15}{8x^\frac{5}{2}y}\big[\frac{\big(3(1-y)-x\big)\tanh^{-1}\sqrt{\frac{x}{1-y}}}{\sqrt{1-y}}-(3-x)\tanh^{-1}\sqrt{x}\big]$\\
\hline
  \hline
$3$ & $\frac{3}{2}$ & $1$ & $4$& $2$ & -$\frac{4}{x^3y}\big[\big(4(1-y)-x\big)\sqrt{\frac{1-x-y}{1-y}}-(4-x)\sqrt{1-x}+4y\big]$\\
\hline
  \hline
$3$ & $2$ & $1$ & -$\frac{1}{2}$& $2$ & $\frac{1}{8y}\big[\frac{4(1-y)^2-48x(1-y)-61x^2}{(1-x-y)^4}-\frac{4-48x-61x^2}{(1-x)^4}-15x^\frac{3}{2}\big(\frac{\big(5(1-y)+2x\big)\arcsin\sqrt{\frac{x}{1-y}}}{(1-x-y)^\frac{9}{2}}-\frac{(5+2x)\arcsin\sqrt{x}}{(1-x)^\frac{9}{2}}\big)\big]$\\
\hline
  \hline
$3$ & $2$ & $1$ & $\frac{1}{2}$& $2$ & $\frac{1}{8y}\big[\frac{4(1-y)+11x}{(1-x-y)^3}-\frac{4+11x}{(1-x)^3}+3\sqrt{x}\big(\frac{\big(3(1-y)+2x\big)\arcsin\sqrt{\frac{x}{1-y}}}{(1-x-y)^\frac{7}{2}}-\frac{(3+2x)\arcsin\sqrt{x}}{(1-x)^\frac{7}{2}}\big)\big]$\\
\hline
  \hline
$3$ &  $2$    & $1$  & $1$& $2$  & $\frac{1}{2y}\big[\frac{1+x-y}{(1-x-y)^3}-\frac{1+x}{(1-x)^3}\big]$ \\
\hline
  \hline
$3$ & $2$ & $1$ & $\frac{3}{2}$& $2$ & $\frac{1}{8y}\big[\frac{3}{(1-x-y)^2}-\frac{3}{(1-x)^2}+\frac{1}{\sqrt{x}}\big(\frac{(1-y+2x)\arcsin\sqrt{\frac{x}{1-y}}}{(1-x-y)^\frac{5}{2}}-\frac{(1+2x)\arcsin\sqrt{x}}{(1-x)^\frac{5}{2}}\big)\big]$\\
\hline
  \hline
\end{tabular*}
\end{center}
\begin{center}
\begin{tabular*}{1.0\textwidth}{@{\extracolsep{\fill}} | l || l | l || l| l|| l | }
  \hline
\textbf{$\sigma$} &  $\alpha_1$    & $\alpha_2$  & $\beta_1$& $\beta_2$  & $F_2(\sigma;\alpha_1,\alpha_2;\beta_1,\beta_2;x,y)$ \\
  \hline\hline
$3$ &  $3$    & $1$  & $4$& $2$  & $\frac{3}{2x^3y}\big[\frac{x^2y}{(1-x)(1-x-y)}+2\big((1-y)\ln\frac{1-x-y}{1-y}-\ln(1-x)\big)\big]$ \\
\hline
  \hline
$3$ &  $\frac{7}{2}$    & $1$  & -$\frac{1}{2}$& $2$  & $\frac{1}{2y}\big[\frac{(1-y)^4-20x(1-y)^3-90x^2(1-y)^2-20x^3(1-y)+x^4}{(1-x-y)^6}-\frac{1-20x-90x^2-20x^3+x^4}{(1-x)^6}\big]$ \\
\hline
  \hline
$3$ &  $\frac{7}{2}$    & $1$  & $\frac{1}{2}$& $2$  & $\frac{1}{10y}\big[\frac{5(1-y)^3+45x(1-y)^2+15x^2(1-y)-x^3}{(1-x-y)^5}-\frac{5+45x+15x^2-x^3}{(1-x)^5}\big]$ \\
\hline
  \hline
$3$ &  $\frac{7}{2}$    & $1$  & $1$& $2$  & $\frac{1}{4y}\big[\frac{(1-y)^\frac{3}{2}\big(2(1-y)+5x\big)}{(1-x-y)^\frac{9}{2}}-\frac{2+5x}{(1-x)^\frac{9}{2}}\big]$ \\
\hline
  \hline
$3$ &  $\frac{7}{2}$    & $1$  & $\frac{3}{2}$& $2$  & $\frac{1}{30y}\big[\frac{15(1-y)^2+10x(1-y)-x^2}{(1-x-y)^4}-\frac{15+10x-x^2}{(1-x)^4}\big]$ \\
\hline
  \hline
$3$ &  $\frac{7}{2}$    & $1$  & $2$& $2$  & $\frac{1}{2y}[(1-y)^\frac{3}{2}(1-x-y)^{-\frac{7}{2}}-(1-x)^{-\frac{7}{2}}]$ \\
\hline
  \hline
$3$ &  $\frac{7}{2}$    & $1$  & $\frac{5}{2}$& $2$  & $\frac{1}{10y}\big[\frac{5(1-y)-x}{(1-x-y)^3}-\frac{5-x}{(1-x)^3}\big]$ \\
\hline
  \hline
\end{tabular*}
\end{center}
\begin{center}
\begin{tabular*}{1.0\textwidth}{@{\extracolsep{\fill}} | l || l | l || l| l|| l | }
  \hline
\textbf{$\sigma$} &  $\alpha_1$    & $\alpha_2$  & $\beta_1$& $\beta_2$  & $F_2(\sigma;\alpha_1,\alpha_2;\beta_1,\beta_2;x,y)$ \\
  \hline\hline
$\frac{7}{2}$ & $2$ & $1$ & $3$& $2$ & $\frac{8}{15x^2y}\big[2\big((1-y)^{-\frac{1}{2}}-1\big)-\frac{2(1-y)-3x}{(1-x-y)^\frac{3}{2}}+\frac{2-3x}{(1-x)^\frac{3}{2}}\big]$\\
\hline
  \hline
$\frac{7}{2}$ & $2$ & $1$ & $\frac{7}{2}$& $2$ & $\frac{1}{x^2y}\big[\frac{3(1-y)-2x}{\sqrt{1-y}(1-x-y)}-\frac{3-2x}{1-x}-3\frac{\tanh^{-1}\sqrt{\frac{x}{1-y}}-\tanh^{-1}\sqrt{x}}{\sqrt{x}}\big]$\\
\hline
  \hline
$\frac{7}{2}$ & $2$ & $1$ & $4$& $2$ & $\frac{16}{5x^3y}\big[\frac{4(1-y)-3x}{\sqrt{1-x-y}}-\frac{4-3x}{\sqrt{1-x}}-\frac{4(1-y)-x}{\sqrt{1-y}}+4-x\big]$\\
\hline
  \hline
$\frac{7}{2}$ & $2$ & $1$ & $5$& $2$ & $\frac{32}{5x^4y}\big[\frac{(\sqrt{1-y}-\sqrt{1-x-y})^4}{\sqrt{1-y}}-(1-\sqrt{1-x})^4\big]$\\
\hline
  \hline
$\frac{7}{2}$ &  $\frac{5}{2}$    & $1$  & -$\frac{1}{2}$& $2$  & $\frac{2}{5y}\big[\frac{(1-y)^3-18x(1-y)^2-72x^2(1-y)-16x^3}{(1-x-y)^\frac{11}{2}}-\frac{1-18x-72x^2-16x^3}{(1-x)^\frac{11}{2}}\big]$ \\
\hline
  \hline
$\frac{7}{2}$ &  $\frac{5}{2}$    & $1$  & $\frac{1}{2}$& $2$  & $\frac{2}{15y}\big[\frac{3(1-y)^2+24x(1-y)+8x^2}{(1-x-y)^\frac{9}{2}}-\frac{3+24x+8x^2}{(1-x)^\frac{9}{2}}\big]$ \\
\hline
  \hline
$\frac{7}{2}$ &  $\frac{5}{2}$    & $1$  & $\frac{3}{2}$& $2$  & $\frac{2}{15y}\big[\frac{3(1-y)+2x}{(1-x-y)^\frac{7}{2}}-\frac{3+2x}{(1-x)^\frac{7}{2}}\big]$ \\
\hline
  \hline
\end{tabular*}
\end{center}
\begin{center}
\begin{tabular*}{1.0\textwidth}{@{\extracolsep{\fill}} | l || l | l || l| l|| l | }
  \hline
\textbf{$\sigma$} &  $\alpha_1$    & $\alpha_2$  & $\beta_1$& $\beta_2$  & $F_2(\sigma;\alpha_1,\alpha_2;\beta_1,\beta_2;x,y)$ \\
  \hline\hline
$4$ &  -$\frac{1}{2}$    & $1$  & $\frac{5}{2}$& $2$  & $\frac{1}{64xy}\big[\frac{1}{\sqrt{x}}\big(\frac{\big((1-y)^2+6x(1-y)-15x^2\big)\tanh^{-1}\sqrt{\frac{x}{1-y}}}{(1-y)^\frac{7}{2}}-(1+6x-15x^2)\tanh^{-1}\sqrt{x}\big)-\frac{(1-y-15x)}{(1-y)^3}+1-15x\big]$\\
\hline
  \hline
$4$ &  -$\frac{1}{2}$    & $1$  & $3$& $2$  & $\frac{1}{3y}[(1-y)^{-\frac{7}{2}}(1-x-y)^\frac{1}{2}-(1-x)^\frac{1}{2}]$ \\
\hline
  \hline
$4$ &  -$\frac{1}{2}$    & $1$  & $\frac{7}{2}$& $2$  & $\frac{5}{384x^2y}\big[\frac{3}{\sqrt{x}}\big(\frac{\big((1-y)^2+2x(1-y)+5x^2\big)(1-x-y)\tanh^{-1}\sqrt{\frac{x}{1-y}}}{(1-y)^\frac{7}{2}}-(1+2x+5x^2)(1-x)\tanh^{-1}\sqrt{x}\big)$\\
~&~&~&~&~&$\quad -\frac{\big(3(1-y)^2+4x(1-y)-15x^2\big)}{(1-y)^3}+3+4x-15x^2\big]$\\
\hline
  \hline
$4$ &  -$\frac{1}{2}$    & $1$  & $4$& $2$  & $\frac{2}{105x^3y}\big[(8+12x+15x^2)(1-x)^\frac{3}{2}-\frac{\big(8(1-y)^2+12x(1-y)+15x^2\big)(1-x-y)^\frac{3}{2}}{(1-y)^\frac{7}{2}}\big]$\\
\hline
  \hline
$4$ & $\frac{1}{2}$ & $1$ & -$\frac{1}{2}$& $2$ & $\frac{1}{3y}\big[\frac{1-y-7x}{(1-x-y)^4}-\frac{1-7x}{(1-x)^4}\big]$\\
\hline
  \hline
$4$ & $\frac{1}{2}$ & $1$ & $1$& $2$ & $\frac{1}{24y}\big[\frac{8(1-y)^2-8x(1-y)+3x^2}{\big((1-y)(1-x-y)\big)^\frac{5}{2}}-\frac{8-8x+3x^2}{(1-x)^\frac{5}{2}}\big]$\\
\hline
  \hline
\end{tabular*}
\end{center}
\begin{center}
\begin{tabular*}{1.0\textwidth}{@{\extracolsep{\fill}} | l || l | l || l| l|| l | }
  \hline
\textbf{$\sigma$} &  $\alpha_1$    & $\alpha_2$  & $\beta_1$& $\beta_2$  & $F_2(\sigma;\alpha_1,\alpha_2;\beta_1,\beta_2;x,y)$ \\
  \hline\hline
$4$ & $2$ & $1$ & $\frac{1}{2}$& $2$ & $\frac{1}{48y}\big[\frac{16(1-y)^2+83x(1-y)+6x^2}{(1-y)(1-x-y)^4}-\frac{16+83x+6x^2}{(1-x)^4}+15\sqrt{x}\big(\frac{\big(3(1-y)+4x\big)\arcsin\sqrt{\frac{x}{1-y}}}{(1-x-y)^\frac{9}{2}}-\frac{(3+4x)\arcsin\sqrt{x}}{(1-x)^\frac{9}{2}}\big)\big]$\\
\hline
  \hline
$4$ & $2$ & $1$ & $1$& $2$ & $\frac{1}{3y}\big[\frac{1-y+2x}{(1-x-y)^4}-\frac{1+2x}{(1-x)^4}\big]$\\
\hline
  \hline
$4$ & $2$ & $1$ & $\frac{3}{2}$& $2$ & $\frac{1}{48y}\big[\frac{13(1-y)+2x}{(1-y)(1-x-y)^3}-\frac{13+2x}{(1-x)^3}+\frac{3}{\sqrt{x}}\big(\frac{(1-y+4x)\arcsin\sqrt{\frac{x}{1-y}}}{(1-x-y)^\frac{7}{2}}-\frac{(1+4x)\arcsin\sqrt{x}}{(1-x)^\frac{7}{2}}\big)\big]$\\
\hline
  \hline
$4$ & $2$ & $1$ & $\frac{5}{2}$& $2$ & $\frac{1}{16y}\big[\frac{1-y+2x}{(1-y)(1-x-y)^2}-\frac{1+2x}{(1-x)^2}-\frac{1}{\sqrt{x}}\big(\frac{(1-y-4x)\arcsin\sqrt{\frac{x}{1-y}}}{(1-x-y)^\frac{5}{2}}-\frac{(1-4x)\arcsin\sqrt{x}}{(1-x)^\frac{5}{2}}\big)\big]$\\
\hline
  \hline
$4$ &  $2$    & $1$  & $3$& $2$  & $\frac{1}{3y}[(1-y)^{-1}(1-x-y)^{-2}-(1-x)^{-2}]$ \\
\hline
  \hline
$4$ & $2$ & $1$ & $\frac{7}{2}$& $2$ & $\frac{5}{16x^2}\big[\frac{2}{1-y}+\frac{1}{(1-x)(1-x-y)}-\frac{1}{\sqrt{x}y}\big(\frac{\big(3(1-y)-4x\big)\arcsin\sqrt{\frac{x}{1-y}}}{(1-x-y)^\frac{3}{2}}-\frac{(3-4x)\arcsin\sqrt{x}}{(1-x)^\frac{3}{2}}\big)\big]$\\
\hline
  \hline
$4$ &  $2$    & $1$  & $4$& $2$  & $\frac{1}{x^2}\big[\frac{1}{1-y}+\frac{1}{(1-x)(1-x-y)}\big]+\frac{2}{x^3y}\ln\big(\frac{1-x-y}{(1-y)(1-x)}\big)$ \\
\hline
  \hline
\end{tabular*}
\end{center}
\begin{center}
\begin{tabular*}{1.0\textwidth}{@{\extracolsep{\fill}} | l || l | l || l| l|| l | }
  \hline
\textbf{$\sigma$} &  $\alpha_1$    & $\alpha_2$  & $\beta_1$& $\beta_2$  & $F_2(\sigma;\alpha_1,\alpha_2;\beta_1,\beta_2;x,y)$ \\
  \hline\hline
$4$ &  $\frac{7}{2}$    & $1$  & $4$& $2$  & $\frac{2}{15x^3y}\big[\frac{\sqrt{1-y}\big(8(1-y)^2-20x(1-y)+15x^2\big)}{(1-x-y)^\frac{5}{2}}-\frac{8-20x+15x^2}{(1-x)^\frac{5}{2}}\big]$ \\
\hline
  \hline
$4$ &  $4$    & $1$  & -$\frac{1}{2}$& $2$  & $\frac{1}{384y}\big[\frac{128(1-y)^4-3968x(1-y)^3-26223x^2(1-y)^2-14702x^3(1-y)-280x^4}{(1-x-y)^7}-\frac{128-3968x-26223x^2-14702x^3-280x^4}{(1-x)^7}$\\
~&~&~&~&~&$\quad -315x^\frac{3}{2}\big(\frac{(1-y)\big(35(1-y)^2+84x(1-y)+24x^2\big)\arcsin\sqrt{\frac{x}{1-y}}}{(1-x-y)^\frac{15}{2}}-\frac{(35+84x+24x^2)\arcsin\sqrt{x}}{(1-x)^\frac{15}{2}}\big)\big]$ \\
\hline
  \hline
$4$ &  $4$    & $1$  & $\frac{1}{2}$& $2$  & $\frac{1}{384y}\big[\frac{128(1-y)^3+1779x(1-y)^2+1518x^2(1-y)+40x^3}{(1-x-y)^6}-\frac{128+1779x+1518x^2+40x^3}{(1-x)^6}$\\
~&~&~&~&~&$\quad +105\sqrt{x}\big(\frac{(1-y)\big(5(1-y)^2+20x(1-y)+8x^2\big)\arcsin\sqrt{\frac{x}{1-y}}}{(1-x-y)^\frac{13}{2}}-\frac{(5+20x+8x^2)\arcsin\sqrt{x}}{(1-x)^\frac{13}{2}}\big)\big]$ \\
\hline
  \hline
$4$ &  $4$    & $1$  & $1$& $2$  & $\frac{1}{3y}\big[\frac{(1-y)\big((1-y)^2+6x(1-y)+3x^2\big)}{(1-x-y)^6}-\frac{1+6x+3x^2}{(1-x)^6}\big]$\\
\hline
  \hline
$4$ &  $4$    & $1$  & $\frac{3}{2}$& $2$  & $\frac{1}{384y}\big[\frac{113(1-y)^2+194x(1-y)+8x^2}{(1-x-y)^5}-\frac{113+194x+8x^2}{(1-x)^5}$\\
\hline
  \hline
\end{tabular*}
\end{center}
\begin{center}
\begin{tabular*}{1.0\textwidth}{@{\extracolsep{\fill}} | l || l | l || l| l|| l | }
  \hline
\textbf{$\sigma$} &  $\alpha_1$    & $\alpha_2$  & $\beta_1$& $\beta_2$  & $F_2(\sigma;\alpha_1,\alpha_2;\beta_1,\beta_2;x,y)$ \\
  \hline\hline
$\frac{9}{2}$ &  $\frac{3}{2}$    & $1$  & $\frac{7}{2}$& $2$  & $\frac{2}{7y}[(1-y)^{-2}(1-x-y)^{-\frac{3}{2}}-(1-x)^{-\frac{3}{2}}]$ \\
\hline
  \hline
$\frac{9}{2}$ & $2$ & $1$ & -$\frac{1}{2}$& $2$ & $\frac{2}{7y}\big[\frac{(1-y)^4-20x(1-y)^3-90x^2(1-y)^2-20x^3(1-y)+x^4}{(1-y)^\frac{3}{2}(1-x-y)^6}-\frac{1-20x-90x^2-20x^3+x^4}{(1-x)^6}\big]$\\
\hline
  \hline
$\frac{9}{2}$ & $2$ & $1$ & $\frac{1}{2}$& $2$ & $\frac{2}{35y}\big[\frac{5(1-y)^3+45x(1-y)^2+15x^2(1-y)-x^3}{(1-y)^\frac{3}{2}(1-x-y)^5}-\frac{5+45x+15x^2-x^3}{(1-x)^5}\big]$\\
\hline
  \hline
$\frac{9}{2}$ & $2$ & $1$ & $1$& $2$ & $\frac{1}{7y}\big[\frac{2(1-y)+5x}{(1-x-y)^\frac{9}{2}}-\frac{2+5x}{(1-x)^\frac{9}{2}}\big]$\\
\hline
  \hline
$\frac{9}{2}$ & $2$ & $1$ & $\frac{3}{2}$& $2$ & $\frac{2}{105y}\big[\frac{15(1-y)^2+10x(1-y)-x^2}{(1-y)^\frac{3}{2}(1-x-y)^4}-\frac{15+10x-x^2}{(1-x)^4}\big]$\\
\hline
  \hline
$\frac{9}{2}$ & $2$ & $1$ & $\frac{5}{2}$& $2$ & $\frac{2}{35y}\big[\frac{5(1-y)-x}{(1-y)^\frac{3}{2}(1-x-y)^3}-\frac{5-x}{(1-x)^3}\big]$\\
\hline
  \hline
$\frac{9}{2}$ & $2$ & $1$ & $3$& $2$ & $\frac{8}{105x^2y}\big[\frac{2}{(1-y)^\frac{3}{2}}-2-\frac{2(1-y)-5x}{(1-x-y)^\frac{5}{2}}+\frac{2-5x}{(1-x)^\frac{5}{2}}\big]$\\
\hline
  \hline
\end{tabular*}
\end{center}
\begin{center}
\begin{tabular*}{1.0\textwidth}{@{\extracolsep{\fill}} | l || l | l || l| l|| l | }
  \hline
\textbf{$\sigma$} &  $\alpha_1$    & $\alpha_2$  & $\beta_1$& $\beta_2$  & $F_2(\sigma;\alpha_1,\alpha_2;\beta_1,\beta_2;x,y)$ \\
  \hline\hline
$5$ &  -$\frac{1}{2}$    & $1$  & $\frac{1}{2}$& $2$  & $\frac{1}{192y}\big[\frac{48(1-y)^3-231x(1-y)^2+280x^2(1-y)-105x^3}{(1-y)^4(1-x-y)^3}-\frac{48-231x+280x^2-105x^3}{(1-x)^3}$\\
~&~&~&~&~&$\quad -105\sqrt{x}(\frac{\tanh^{-1}\sqrt{\frac{x}{1-y}}}{(1-y)^\frac{9}{2}}-\tanh^{-1}\sqrt{x})]$\\
\hline
  \hline
$5$ &  -$\frac{1}{2}$    & $1$  & $1$& $2$  & $\frac{1}{64y}\big[\frac{16(1-y)^3-72x(1-y)^2+90x^2(1-y)-35x^3}{(1-y)^\frac{9}{2}(1-x-y)^\frac{5}{2}}-\frac{16-72x+90x^2-35x^3}{(1-x)^\frac{5}{2}}\big]$\\
\hline
  \hline
$5$ &  -$\frac{1}{2}$    & $1$  & $\frac{3}{2}$& $2$  & $\frac{1}{384y}\big[\frac{81(1-y)^2-190x(1-y)+105x^2}{(1-y)^4(1-x-y)^2}-\frac{81-190x+105x^2}{(1-x)^2}+\frac{15}{\sqrt{x}}\big(\frac{(1-y-7x)\tanh^{-1}\sqrt{\frac{x}{1-y}}}{(1-y)^\frac{9}{2}}-(1-7x)\tanh^{-1}\sqrt{x}\big)\big]$\\
\hline
  \hline
$5$ &  -$\frac{1}{2}$    & $1$  & $2$& $2$  & $\frac{1}{96y}\big[\frac{24(1-y)^2-60x(1-y)+35x^2}{(1-y)^\frac{9}{2}(1-x-y)^\frac{3}{2}}-\frac{24-60x+35x^2}{(1-x)^\frac{3}{2}}\big]$\\
\hline
  \hline
$5$ &  -$\frac{1}{2}$    & $1$  & $\frac{5}{2}$& $2$  & $\frac{1}{512xy}\big[\frac{3}{\sqrt{x}}\big(\frac{\big((1-y)^2+10x(1-y)-35x^2\big)\tanh^{-1}\sqrt{\frac{x}{1-y}}}{(1-y)^\frac{9}{2}}-(1+10x-35x^2)\tanh^{-1}\sqrt{x}\big)$\\
~&~&~&~&~&$\quad -\frac{3(1-y)^2+100x(1-y)+105x^2}{(1-y)^4(1-x-y)}+\frac{3+100x+105x^2}{1-x}\big]$\\
\hline
  \hline
\end{tabular*}
\end{center}
\begin{center}
\begin{tabular*}{1.0\textwidth}{@{\extracolsep{\fill}} | l || l | l || l| l|| l | }
  \hline
\textbf{$\sigma$} &  $\alpha_1$    & $\alpha_2$  & $\beta_1$& $\beta_2$  & $F_2(\sigma;\alpha_1,\alpha_2;\beta_1,\beta_2;x,y)$ \\
  \hline\hline
$5$ & $\frac{3}{2}$ & $1$ & $2$& $2$ & $\frac{1}{32y}\big[\frac{8(1-y)^2-4x(1-y)+x^2}{(1-y)^\frac{5}{2}(1-x-y)^\frac{7}{2}}-\frac{8-4x+x^2}{(1-x)^\frac{7}{2}}\big]$\\
\hline
  \hline
$5$ & $\frac{3}{2}$ & $1$ & $\frac{5}{2}$& $2$ & $\frac{1}{64xy}\big[\frac{3(1-y)^2+8x(1-y)-3x^2}{(1-y)^2(1-x-y)^3}-\frac{3+8x-3x^2}{(1-x)^3}-\frac{3}{\sqrt{x}}\big(\frac{\tanh^{-1}\sqrt{\frac{x}{1-y}}}{(1-y)^\frac{5}{2}}-\tanh^{-1}\sqrt{x}\big)\big]$\\
\hline
  \hline
$5$ & $\frac{3}{2}$ & $1$ & $3$& $2$ & $\frac{1}{8y}\big[\frac{2(1-y)-x}{\big((1-y)(1-x-y)\big)^\frac{5}{2}}-\frac{2-x}{(1-x)^\frac{5}{2}}\big]$\\
\hline
  \hline
$5$ & $\frac{3}{2}$ & $1$ & $\frac{7}{2}$& $2$ & $\frac{5}{128x^2y}\big[\frac{3(1-y)^2-2x(1-y)+3x^2}{\big((1-y)(1-x-y)\big)^2}-\frac{3-2x+3x^2}{(1-x)^2}-\frac{3}{\sqrt{x}}\big(\frac{(1+x-y)\tanh^{-1}\sqrt{\frac{x}{1-y}}}{(1-y)^\frac{5}{2}}-(1+x)\tanh^{-1}\sqrt{x}\big)\big]$\\
\hline
  \hline
$5$ &  $\frac{3}{2}$    & $1$  & $4$& $2$  & $\frac{1}{4y}[(1-y)^{-\frac{5}{2}}(1-x-y)^{-\frac{3}{2}}-(1-x)^{-\frac{3}{2}}]$ \\
\hline
  \hline
$5$ & $2$ & $1$ & -$\frac{1}{2}$& $2$ & $\frac{1}{128y}\big[\frac{32(1-y)^4-704x(1-y)^3-2553x^2(1-y)^2-260x^3(1-y)+20x^4}{(1-y)^2(1-x-y)^6}-\frac{32-704x-2553x^2-260x^3+20x^4}{(1-x)^6}$\\
~&~&~&~&~&$\quad -315x^\frac{3}{2}\big(\frac{\big(5(1-y)+6x\big)\arcsin\sqrt{\frac{x}{1-y}}}{(1-x-y)^\frac{13}{2}}-\frac{(5+6x)\arcsin\sqrt{x}}{(1-x)^\frac{13}{2}}\big)\big]$\\
\hline
  \hline
\end{tabular*}
\end{center}
\begin{center}
\begin{tabular*}{1.0\textwidth}{@{\extracolsep{\fill}} | l || l | l || l| l|| l | }
  \hline
\textbf{$\sigma$} &  $\alpha_1$    & $\alpha_2$  & $\beta_1$& $\beta_2$  & $F_2(\sigma;\alpha_1,\alpha_2;\beta_1,\beta_2;x,y)$ \\
  \hline\hline
$5$ & $3$ & $1$ & $\frac{7}{2}$& $2$ & $\frac{5}{512x^2y}\big[\frac{3}{\sqrt{x}}\big(\frac{\big((1-y)^2-4x(1-y)+8x^2\big)\arcsin\sqrt{\frac{x}{1-y}}}{(1-x-y)^\frac{7}{2}}-\frac{(1-4x+8x^2)\arcsin\sqrt{x}}{(1-x)^\frac{7}{2}}\big)$\\
~&~&~&~&~&$\quad -\frac{3(1-y)^2-10x(1-y)-8x^2}{(1-y)(1-x-y)^3}+\frac{3+94x+8x^2}{(1-x)^3}\big]$\\
\hline
  \hline
$5$ &  $3$    & $1$  & $4$& $2$  & $\frac{1}{4y}[(1-y)^{-1}(1-x-y)^{-3}-(1-x)^{-3}]$ \\
\hline
  \hline
$5$ & $\frac{7}{2}$ & $1$ & -$\frac{1}{2}$& $2$ & $\frac{1}{4y}\big[\frac{(1-y)^4-36(1-y)^3-378x^2(1-y)^2-420x^3(1-y)-63x^4}{(1-x-y)^8}-\frac{1-36-378x^2-420x^3-63x^4}{(1-x)^8}\big]$\\
\hline
  \hline
$5$ & $\frac{7}{2}$ & $1$ & $\frac{1}{2}$& $2$ & $\frac{1}{4y}\big[\frac{(1-y)^3+21x(1-y)^2+35x^2(1-y)+7x^3}{(1-x-y)^7}-\frac{1+21x+35x^2+7x^3}{(1-x)^7}\big]$\\
\hline
  \hline
$5$ & $\frac{7}{2}$ & $1$ & $1$& $2$ & $\frac{1}{64y}\big[\frac{16(1-y)^3+120x(1-y)^2+90x^2(1-y)+5x^3}{\sqrt{1-y}(1-x-y)^\frac{13}{2}}-\frac{16+120x+90x^2+5x^3}{(1-x)^\frac{13}{2}}\big]$\\
\hline
  \hline
$5$ & $\frac{7}{2}$ & $1$ & $\frac{3}{2}$& $2$ & $\frac{1}{12y}\big[\frac{3(1-y)^2+10x(1-y)+3x}{(1-x-y)^6}-\frac{3+10x+3x^2}{(1-x)^6}\big]$\\
\hline
  \hline
\end{tabular*}
\end{center}

\section{Representations of Appell hypergeometric function $F_2$}

\begin{center}
\begin{tabular*}{1.0\textwidth}{@{\extracolsep{\fill}} | l || l | l || l| l|| l | }
  \hline
\textbf{$\sigma$} &  $\alpha_1$    & $\alpha_2$  & $\beta_1$& $\beta_2$  & $F_2(\sigma;\alpha_1,\alpha_2;\beta_1,\beta_2;x,y)$ \\
  \hline\hline
$\sigma$ & $\alpha_1$ & $1$ & $\beta_1$& $2$ & $(1-x)^{-\sigma}F_2\big(\sigma;\beta_1-\alpha_1,1;\beta_1,2;\frac{x}{x-1},\frac{y}{1-x}\big), (\sigma\neq{0})$\\
\hline
  \hline
$a+1$ &  $a-\frac{1}{2}$    & $1$  & $2a$& $2$  & $\frac{1}{ay}\big[\frac{1}{\sqrt{1-y}}\big(\frac{\sqrt{1-y}+\sqrt{1-x-y}}{2}\big)^{1-2a}-\big(\frac{1+\sqrt{1-x}}{2}\big)^{1-2a}\big]$ \\
\hline
  \hline
$a+1$ &  $a+\frac{1}{2}$    & $1$  & $2a$& $2$  & $\frac{1}{ay}\big[\frac{1}{\sqrt{1-x-y}}\big(\frac{2}{\sqrt{1-y}+\sqrt{1-x-y}}\big)^{2a-1}-\frac{1}{\sqrt{1-x}}\big(\frac{2}{1+\sqrt{1-x}}\big)^{2a-1}\big]$ \\
\hline
  \hline
$a+1$ &  $a+\frac{1}{2}$    & $1$  & $2a+1$& $2$  & $\frac{1}{ay}\big[\big(\frac{2}{\sqrt{1-y}+\sqrt{1-x-y}}\big)^{2a}-\big(\frac{2}{1+\sqrt{1-x}}\big)^{2a}\big],\qquad\qquad(a\neq{-\frac{1}{2}, -1})$ \\
\hline
  \hline
$a+1$ &  $a+\frac{1}{2}$    & $1$  & $\frac{1}{2}$& $2$  & $\frac{(\sqrt{1-y}+\sqrt{x})^{-2a}+(\sqrt{1-y}-\sqrt{x})^{-2a}-(1+\sqrt{x})^{-2a}-(1-\sqrt{x})^{-2a}}{2ay}$ \\
\hline
  \hline
$a+1$ & $a+\frac{1}{2}$ & $1$ & $\frac{3}{2}$& $2$ & $\frac{1}{2a(2a-1)\sqrt{x}y}\big[(\sqrt{1-y}-\sqrt{x})^{1-2a}-(1-\sqrt{x})^{1-2a}-(\sqrt{1-y}+\sqrt{x})^{1-2a}+(1+\sqrt{x})^{1-2a}\big]$\\
\hline
  \hline
$a+1$ &  $\frac{a}{2}+1$    & $1$  & $\frac{a}{2}$& $2$  & $\frac{1}{ay}\big[\frac{1-y+x}{(1-x-y)^{a+1}}-\frac{1+x}{(1-x)^{a+1}}\big]$ \\
\hline
  \hline
\end{tabular*}
\end{center}
\begin{center}
\begin{tabular*}{1.0\textwidth}{@{\extracolsep{\fill}} | l || l | l || l| l|| l | }
  \hline
\textbf{$\sigma$} &  $\alpha_1$    & $\alpha_2$  & $\beta_1$& $\beta_2$  & $F_2(\sigma;\alpha_1,\alpha_2;\beta_1,\beta_2;x,y)$ \\
  \hline\hline
$2$ & $1$ & $1$ & $m$& $2$ & $\frac{(m-1)x}{y}\big[\sum_{k=2}^{m-1}\frac{(x+y-1)^{k-2}-(x-1)^{k-2}}{(m-k)x^k}$\\
~&~&~&~&~&$\quad -x^{-m}\big((x+y-1)^{m-2}\ln\big(\frac{1-x-y}{1-y}\big)-(x-1)^{m-2}\ln(1-x)\big)\big]$\\
\hline
  \hline
$2$ & $\frac{1}{2}$ & $1$ & $m+\frac{1}{2}$& $2$ & $\frac{(\frac{1}{2})_m}{(m-1)!x^my}\big[2\sqrt{x}\big(\frac{(x+y-1)^{m-1}\tanh^{-1}\sqrt{\frac{x}{1-y}}}{\sqrt{1-y}}-(x-1)^{m-1}\tanh^{-1}\sqrt{x}\big)$\\
~&~&~&~&~&$\quad +\sum_{k=1}^{m-2}\frac{(k-1)!x^k}{(\frac{1}{2})_k}\big((x+y-1)^{m-k-1}-(x-1)^{m-k-1}\big)$\\
\hline
  \hline
$2$ & $1$ & $1$ & $m+\frac{1}{2}$& $2$ & $\frac{2m-1}{x^my}\big[(-1)^{m-1}\sqrt{x}\big(\frac{\arcsin\sqrt{\frac{x}{1-y}}}{(1-x-y)^{\frac{3}{2}-m}}-\frac{\arcsin\sqrt{x}}{(1-x)^{\frac{3}{2}-m}}\big)$\\
~&~&~&~&~&$\quad +\sum_{k=1}^{m-1}\frac{x^k}{2k-1}\big((x+y-1)^{m-k-1}-(x-1)^{m-k-1}\big)\big],\qquad\qquad(m=1,2,3,...)$\\
\hline
  \hline
$2$ & -$n$ & $1$ & $m$& $2$ & $\frac{n!(-x)^{1-m}}{(m)_ny}\big[\frac{(1-x-y)^{m+n-1}}{(1-y)^{n+1}}-(1-x)^{m+n-1}-\sum_{k=0}^{m-3}\frac{(1-m-n)_kx^k}{k!}\big((1-y)^{m-k-2}+1\big)\big]$\\
\hline
  \hline
\end{tabular*}
\end{center}
\begin{center}
\begin{tabular*}{1.0\textwidth}{@{\extracolsep{\fill}} | l || l | l || l| l|| l | }
  \hline
\textbf{$\sigma$} &  $\alpha_1$    & $\alpha_2$  & $\beta_1$& $\beta_2$  & $F_2(\sigma;\alpha_1,\alpha_2;\beta_1,\beta_2;x,y)$ \\
  \hline\hline
$1-n$ & $b$ & $1$ & $c$& $2$ & -$\frac{1}{ny}\sum_{k=0}^{n-1}\frac{(-n)_k(b)_kx^k}{(c)_kk!}\big((1-y)^{n-k}-1\big)$\\
~&~&~&~&~&$\quad = -\frac{x^{1-c}}{n(c)_ny}\big[\frac{(1-x-y)^{n+c-b}}{(1-y)^{1-b}}\big(\frac{d}{d(\frac{x}{1-y})}\big)^n[(\frac{x}{1-y})^{n+c-1}(1-\frac{x}{1-y})^{b-c}]$\\
~&~&~&~&~&$\quad\quad\quad -(1-x)^{n+c-b}\big(\frac{d}{dx}\big)^n[x^{n+c-1}(1-x)^{b-c}]\big],\qquad(n\neq{0})$\\
\hline
  \hline
\end{tabular*}
\end{center}

\bigskip
\section*{Acknowledgments}
\medskip
\noindent The present study was supported, in part, by the {\it Natural Sciences and Engineering Research Council of Canada} under Grant GP249507 (NS).  
\medskip



\end{document}